\documentclass[12pt]{iopart}

\usepackage{iopams}  
\usepackage{tikz}

\usetikzlibrary{arrows}
\usetikzlibrary{decorations.pathmorphing}
\usetikzlibrary{decorations.markings}
\usetikzlibrary{arrows.meta}
\usepackage{braket}
\usepackage{setspace}
\usepackage{hyperref}
\usepackage{bm}
\usepackage{graphicx}
\usepackage{placeins}
\usepackage{changepage}
\usepackage{titlepic}
\usepackage[sort&compress,numbers]{natbib}
\usepackage{xcolor}
\usepackage{amssymb}
\usepackage{csquotes}

\newcommand{\tcred}[1]{\textcolor[rgb]{0.6,0,0}{#1}}

\begin{document}

\title[Coherent trapping in small quantum networks]{
Coherent trapping in small quantum networks}

\author{T.J.~Pope,$^1$, J.~Rajendran$^{2,3}$, A.~Ridolfo,$^{2,3}$, E.~Paladino$^{2,3,4}$ , F.M.D. Pellegrino$^{2,3}$, and G.~Falci$^{2,3,4}$}

\address{$^1$ School of Natural and Environmental Sciences, Newcastle University, Newcastle upon Tyne NE1 7RU} 
\address{$^2$ Dipartimento di Fisica e Astronomia \enquote{Ettore Majorana}, Via S. Sofia 64, Ctanaia (I)}
\address{$^3$ Istituto Nazionale di Fisica Nucleare, Sezione di Catania, Via S. Sofia 64, Ctanaia (I) }
\address{$^4$ Consiglio Nazionale delle Ricerche, Istituto di Microelettronica e Microsistemi, UoS Universit\`a, Via S. Sofia 64, Ctanaia (I)}   
\ead{gfalci@dmfci.unict.it}

\begin{abstract}
We consider a three-node fully connected network (Delta network) showing that  
a coherent population trapping phenomenon occurs, generalizing results for  
the Lambda network known to support a dark state. Transport in such structures 
provides signatures of detrapping, which 
can be triggered by external controls. In the presence of an environment it turns out to
be sensitive to its Markovianity. Adiabatic modulation of the system's parameters may yield coherent population transfer, analogous to the stimulated Raman adiabatic passage phenomenon. Robustness of this protocol against non-adiabatic transitions is studied. Coherent nanostructures where these phenomena are relevant for quantum transport and quantum protocols are suggested.
\end{abstract}

%
%
%
%
%

\section{Introduction}
\begin{figure}[!t]
\centering
  {
	\begin{minipage}[c][1\width]{
	   0.3\textwidth}
	   \centering
	   \begin{tikzpicture}[
      scale=0.6,
      level/.style={ultra thick},
      decay/.style={->,decorate,decoration={snake,amplitude=.9mm,segment length=2mm,post length=1mm}, draw=black},
      virtual/.style={thick,densely dashed},
      trans/.style={thin,<->,shorten >=2pt,shorten <=2pt,>=stealth, draw=blue},
      tran2S/.style={thin,<->,shorten >=2pt,shorten <=2pt,>=stealth, draw=red},
      classical/.style={thin,double,<->,shorten >=4pt,shorten <=4pt,>=stealth}
    ]
    \draw[level] (2.5cm,-11em) -- (0.5cm,-11em) node[midway,below] {$\ket{0}$};
    \draw[level] (2cm,6em) -- (4cm,6em) node[midway,above] {$\ket{2}$};
    \draw[level] (4cm,-3em) -- (6cm,-3em) node[midway,below] {$\ket{1}$};
    \draw[level] (7cm,-7em) -- (9cm,-7em) node[midway,below] {$\ket{S}$};
    \draw[trans] (1cm,-11em) -- (2.5cm,6em) node[midway,left] {$\Omega_{P}$};
    \draw[trans] (3.5cm,6em) -- (5cm,-3em) node[midway,left] {$\Omega_{S}$};
    \draw[tran2S] (4.5cm,-3em) -- (2cm,-11em) node[midway,left] {$\Omega_{0}$};
    \draw[decay] (4cm,6em) -- (8cm,-6.8em);
    \draw (7.5cm,-2.1em) node {$\Gamma_{2S}$};
    \end{tikzpicture} 
	\end{minipage}}
 \hspace{30pt}
    {
	\begin{minipage}[c][1\width]{
	   0.3\textwidth}
	   \centering
	   \begin{tikzpicture}[
      scale=0.6,
      level/.style={ultra thick},
      decay/.style={->,decorate,decoration={snake,amplitude=.9mm,segment length=2mm,post length=1mm}, draw=black},
      virtual/.style={thick,densely dashed},
      trans/.style={thin,<->,shorten >=2pt,shorten <=2pt,>=stealth, draw=blue},
      tran2S/.style={thin,<->,shorten >=2pt,shorten <=2pt,>=stealth, draw=red},
      classical/.style={thin,double,<->,shorten >=4pt,shorten <=4pt,>=stealth}
    ]
    \draw[level] (2.5cm,-11em) -- (0.5cm,-11em) node[midway,below] {$\ket{D}$};
    \draw[level] (2cm,-3em) -- (4cm,-3em) node[near end,below] {$\ket{B}$};
    \draw[level] (4cm,6em) -- (6cm,6em) node[midway,below] {$\ket{2}$};
    \draw[level] (7cm,-8em) -- (9cm,-8em) node[midway,below] {$\ket{S}$};
    \draw[trans] (3.5cm,-3em) -- (4.5cm,6em) node[midway,right] {$\Omega_{\mathrm{RMS}}$};
    \draw[tran2S] (1.5cm,-11em) -- (2.5cm,-3em) node[midway,right] {$\delta$};
    \draw[decay] (5.5cm,6em) -- (8cm,-7.8em);
    \draw (7.55cm,-2.1em) node {$\Gamma_{S}$};
    \end{tikzpicture}
	\end{minipage}}
\caption{Left panel schematic of the $\Delta$ system consisting of three levels and three couplings.
One of the levels may dacay to a sink with a rate $\Gamma_{2S}$. This network may admit a stable \enquote{trapped} eigenstate $\ket{D}$, the population being confined in the subspace $\{\ket{0},\ket{1}\}$, if the condition $\bar{\delta}=0$ is fullfilled (see text). 
This is an effective model for a simple quantum network, or for a three-level atom driven by three coherent fields. In the $\Lambda$ system one of the couplings vanishes, $\Omega_0=0$. 
Right panel: the same network in a \enquote{dressed} basis, emphasizing that detrapping occurs 
only for $\bar{\delta} \neq 0$.  
\label{fig:delta_scheme}}
\end{figure}
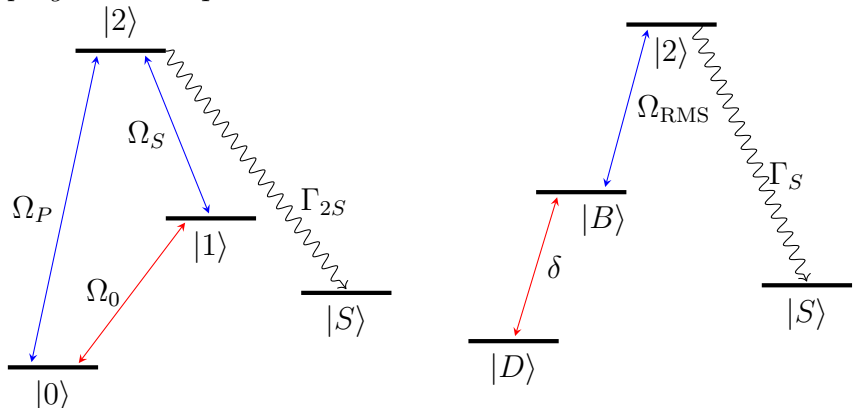

Coherent trapping is a well known phenomenon in atomic physics, occurring when various ground states are coupled to a common upper level by multiple resonant laser beams~\cite{ka:76-arimondo-nuovocimento,ka:78-graystroud-optlett-cohertrap}, the simplest instance being a three-level system called Lambda ($\Lambda$) network (see Fig.~\ref{fig:delta_scheme}).
As a result population is trapped in a subspace, the upper level being never populated. Since this prevents radiative decay the trapped state is often called dark state. Adiabatic manipulations of the dark state may allow faithful and robust population transfer by techniques as the stimulated emission Raman adiabatic passage (STIRAP)~\cite{kr:17-vitanovbergmann-rmp} or superadiabatic extensions~\cite{ka:14-giannelliarimondo-pra-superad}, and are nowadays of interest for applications in solid-state quantum information~\cite{ka:205-liunori-prl-adiabaticpassage,ka:09-siebrafalci-prb,ka:16-vepsalainen-photonics-squtrit,kr:11-younori-nature-artificialatoms,ka:13-zhou-prl-deltasysphotonrouting,ka:19-falci-scirep-usc,ka:18-ridolfofalci-epj-usc}. 

In this work we study a Delta ($\Delta$) systems, a three-site quantum network depicted in  Fig.~\ref{fig:delta_scheme}. 
Each site defines a basis state $\{\ket{0},\ket{1},\ket{2}\}$. 
We will show that under suitable conditions also this more general network admits a stable state, 
trapped in the subspace $\{\ket{0},\ket{1}\}$. The Hamiltonian of the 
$\Delta$ systems is written as 
\begin{eqnarray}
H=\left(
\hskip-2pt
\begin{array}{ccc}
0&\Omega_{0}&\Omega_p
\\
\Omega_{0}^* &\delta &\Omega_s\\
\Omega_p & \Omega_s & \delta_p
\end{array}
\hskip-2pt
\right)
\label{eq:hamiltonian}
\end{eqnarray}
where we have chosen the arbitrary phases of the basis states in such a way that the two off-diagonal elements $\Omega_p$ and $\Omega_s$ are real.
Here we use the notation of quantum optics where Eq.(\ref{eq:hamiltonian}) is the Hamiltonian of a multilevel atom driven by three nearly resonant lasers expressed in a multiple rotating frame, in the rotating wave approximation. Diagonal elements represent detunings of the fields while off-diagonal elements are the Rabi energies, related to the amplitudes (see \ref{sec:stirap}) of the driving fields. When the system is driven by only two lasers special configurations are obtained:
the $\Lambda$ scheme for $\Omega_0=0$, the ladder scheme ($\Omega_p=0$) and the Vee scheme ($\Omega_s=0$). 

Seen as a Hamiltonian in the laboratory frame Eq.(\ref{eq:hamiltonian}) describes a quantum network where for instance three localized states with on-site energies $\delta$ and $\delta_p$ (relative to $\ket{0}$) and with $\Omega_i$ being the tunneling amplitudes connecting the sites. The physics of population trapping also plays an important in solid-state quantum networks~\cite{ka:11-mulkenblumen-physrept,ka:15-bianconi-epl-revnetw} and has been invoked to explain peculiar quantum transport properties in nanostructures, from architectures of quantum dots~\cite{ka:14-plenio-njp-tripleqdot} to light-harvesting systems in biological complexes~\cite{kr:213-lambertnori-nphys-qbiol,kb:14-mohseni-qbiol}. 

We will show in Sec.~\ref{sec:trapped} that also in the $\Delta$ system an \enquote{unconventional} trapping phenomenon occurs. We discuss in Sec.~\ref{sec:qunet} the implications for transport in solid-state quantum networks and in Sec.~\ref{sec:stirap} applications to the dynamics of driven three-level atoms, who also provide insight on the robustness of unconventional trapping. 

\section{Trapped states}
\label{sec:trapped}
We now look for trapped states of the network, namely if eigenstates of the form 
$$
\ket{D} = c_0 \ket{0} + c_1 \ket{1}
$$
exist. The probability of detecting the system in state $\ket{2}$ vanishes, $\langle{2}|{D}\rangle=0$, therefore the system is trapped in a one-dimensional subspace, despite the non vanishing transition amplitudes. In the quantum optical version the system is trapped in $\ket{D}$ despite the fields tend to trigger transitions to the state $\ket{2}$. Since this latter often decays radiatively, when the system is trapped no fluorescence is observed, therefore $\ket{D}$ is called a dark state.

By imposing that $\ket{D}$ is an eigenstate  we find the equations
$$
\left\{
\begin{array}{l}
- z_D c_0 + \Omega_0 c_1 = 0
\\
\Omega_0^* c_0 + (\delta - z_D) c_1 =0  
\\
\Omega_p c_0 + \Omega_s c_1 = 0
\end{array}
\right.
$$ 
$z_D$ being the relative eigenvalue $z$. 
The last equation gives $c_1 = - (\Omega_p/\Omega_s) c_0$, yielding the form of the trapped state
\begin{equation}
\label{eq:trapped}
\ket{D} = {\Omega_s \ket{0} - \Omega_p \ket{1}\over \Omega_{RMS}} =: 
\cos \theta \ket{0} - \sin \theta \ket{1}
\end{equation}
where $\Omega_{RMS}=\sqrt{\Omega_p^2 + \Omega_s^2}$ and the mixing angle is given by 
$\tan \theta = \Omega_p/\Omega_s$. This trapped state is an eigenstate only if the other two 
equations are satisfied, namely 
\begin{eqnarray}
\label{eq:dark-condition}
\delta &=  {\Omega_0^* \Omega_s^2 - \Omega_0 \Omega_p^2 \over \Omega_p \Omega_s} =: \delta_0
\\
z&= - {\Omega_0 \Omega_p \over \Omega_s} =: z_D
\label{eq:dark-eigenvalue}
\end{eqnarray}
Notice that besides $\Omega_0$, also $\delta$ and $\delta_p$ may be complex. For simplicity here 
we focus on the case of real matrix elements. The network can be digonalized and we find 
for the eigenvalues the compact form
\begin{equation}
\label{eq:eigenvalues}
z_\pm = {1 \over 2} \,
\Big[\delta_p + {\Omega_0 \Omega_s \over \Omega_p} \pm \Omega_{AT}
\Big]
\end{equation}
where a quantity analogous to the Autler-Townes splitting appears
\begin{equation}
\label{eq:parameters1}
\Omega_{AT} = \sqrt{\tilde{\delta}_p^2 + 4 \Omega_{RMS}^2} \qquad ; \qquad
\tilde{\delta}_p =  \delta_p - {\Omega_0 \Omega_s \over \Omega_p}
\end{equation}
Eigenvectors can be conveniently written as 
\begin{equation}
\label{eq:eigenvectors}
\ket{+} =  \sin \Phi \, \ket{B} + \cos \Phi \,\ket{2} \quad ; \quad
\ket{-} =  \cos \Phi \, \ket{B} - \sin \Phi \,\ket{2}
\end{equation}
where $\ket{B} = \sin \theta \, \ket{0} + \cos \theta \,\ket{1}$ is called bright state, and 
forms a basis together with $\ket{D}$ and $\ket{2}$. Finally the mixing angle is given by
\begin{equation}
\label{eq:parameter2}
\tan 2 \Phi = {2 \Omega_{RMS} \over \tilde{\delta}_p}
\end{equation}
A well known special solution is obtained for $\Omega_0=0$, which describes a Lambda system. 
The dark state is obtained at zero two-photon detuning, $\delta =0$, and the eigenvalue is $z_D=0$, independent on all the other parameters. This latter is a strong indication of the robustness of the 
dark state for fluctuations of the parameters. We will see that robustness is kept also for $\Omega_0 \neq 0$, despite of the fact that $z_D \neq 0$.
Another special case is the symmetric Delta system, where $\Omega_s = \Omega_p$, the dark state being obtained for $\delta=0$, with eigenvalue $z_d=- \Omega_0$. 

Notice that eigenvalues Eq.~(\ref{eq:eigenvalues}) and eigenvectors Eq.~(\ref{eq:eigenvectors}) have the same structure known for Lambda systems~\cite{kr:17-vitanovbergmann-rmp}, with redefined parameters given by Eqs.(\ref{eq:parameters1},\ref{eq:parameter2}).

\section{Trapped states in nanostructures and detrapping}
\label{sec:qunet}
\begin{figure}[t!] 
 \centering
\includegraphics[width=0.47\textwidth]{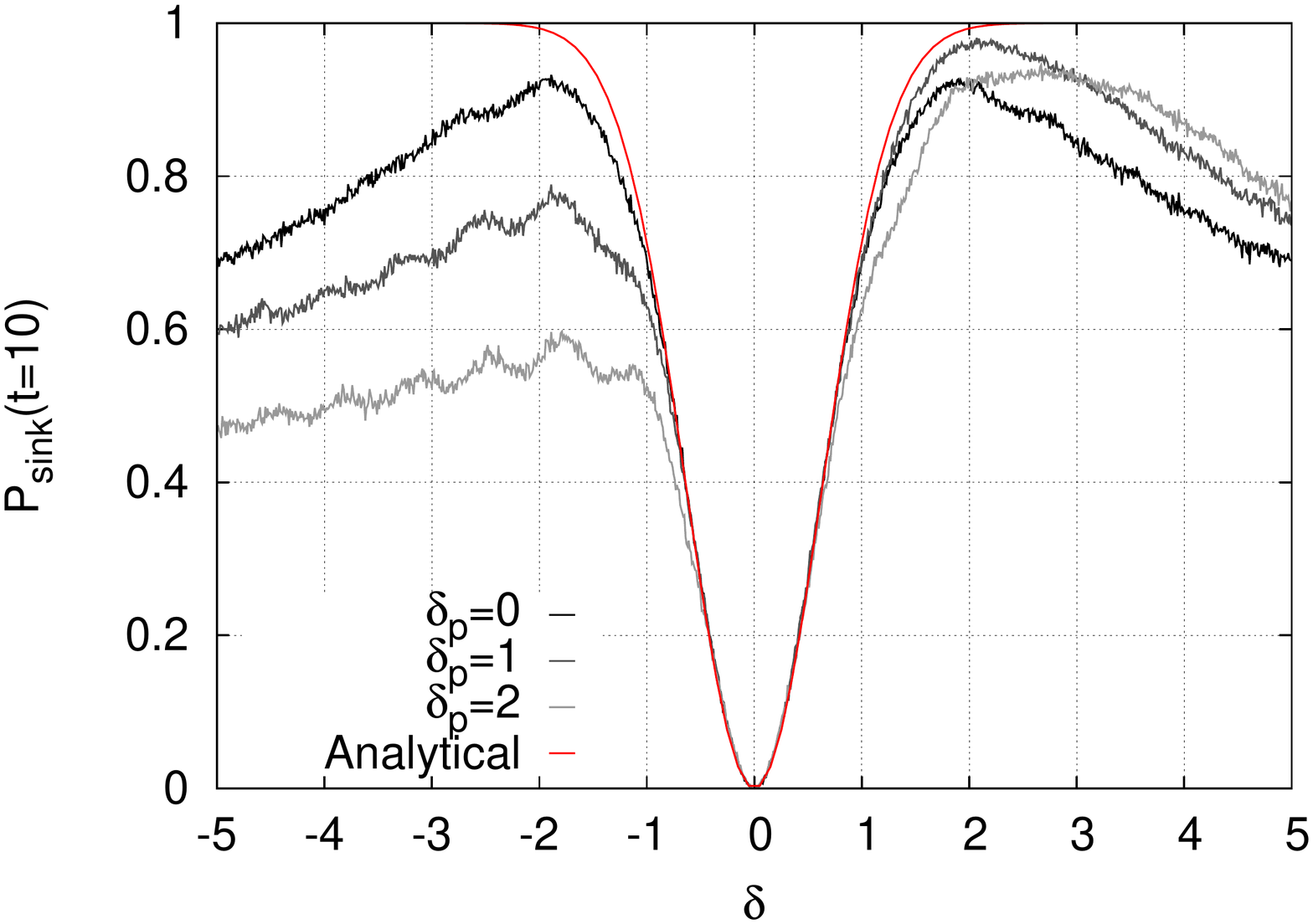}
\includegraphics[width=0.47\textwidth]{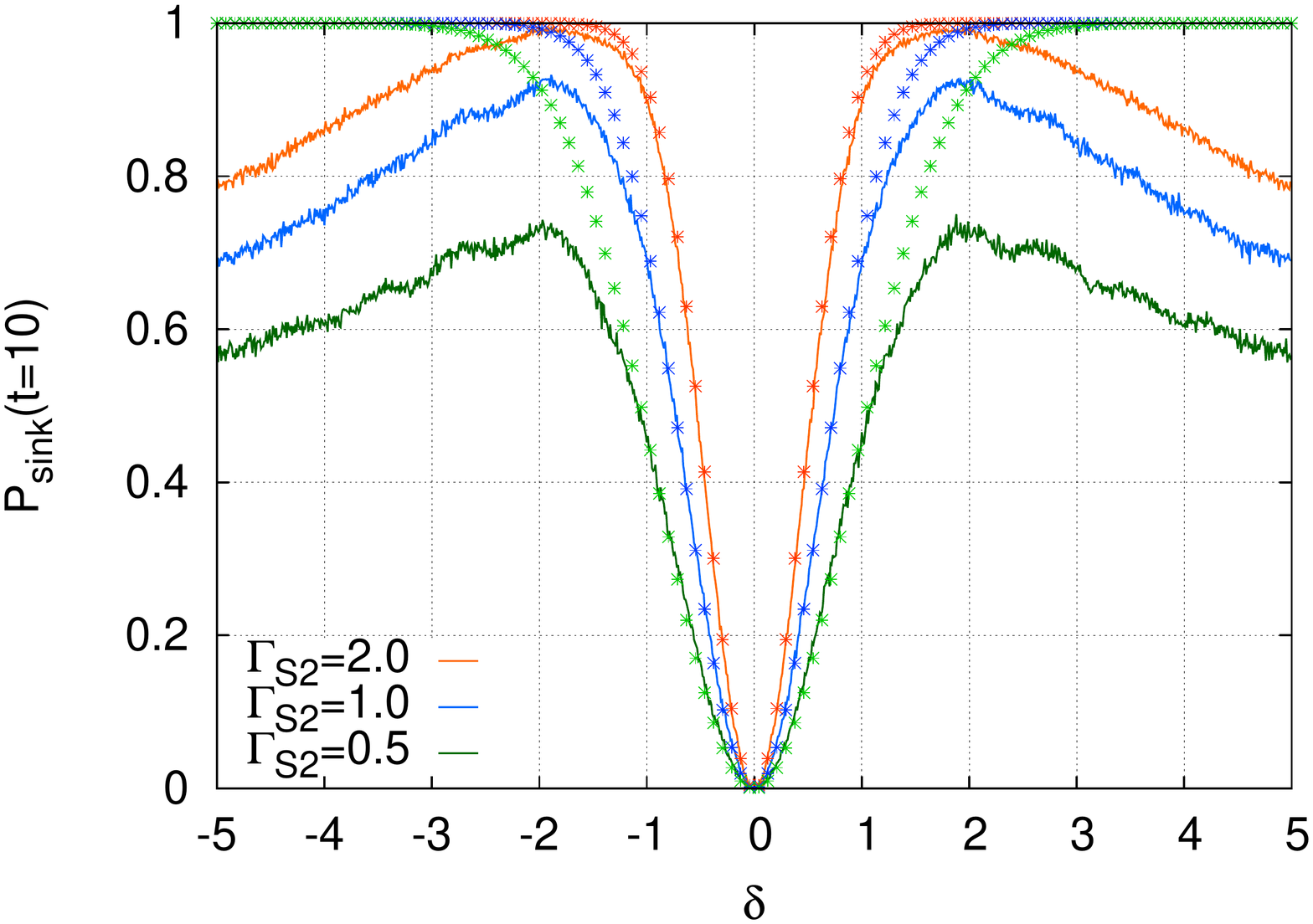}
\caption{Numerically calculated final sink population for the $\Lambda$ network. Here 
$\Omega_s = \Omega_p = 1$ thus $\Omega_{RMS}= \sqrt{2}$. 
Left: sink population for for a range of $\delta$ and $\delta_{p}$, with $\Gamma_{2S}=1$, 
compared to the analytical prediction (\ref{eq:sinkpop}).
Right: Numerically calculated final sink population for the $\Lambda$ network for a range of $\delta$ and $\Gamma_{2S}$, where $\Omega=1$ and $\delta_{p}=\Omega_{0}=0$, compared to the analytical prediction for sink. 
\label{fig:lambda_lightbasis}}
\end{figure}
\subsection{Transport in small quantum networks}
Dark states in Lambda and Delta systems play a crucial role in the dynamics of quantum networks.  
The Hamiltonian (\ref{eq:hamiltonian}) describes three sites where an electron or an exciton 
can be trapped, the off-diagonal elements describing tunneling. We also allow decay from the site 
$\ket{2}$ outside of the system. 
To understand the dynamics it is convenient to represent the Hamiltonian in the basis $\{\ket{D},\ket{B},\ket{2}\}$   
\begin{equation}
\label{eq:hamil-bright}
H =\left( \hskip-2pt \begin{array}{ccc}
- {\Omega_0 \Omega_p \over \Omega_s}&0&0
\\
0 &{\Omega_0 \Omega_s \over \Omega_p} & \Omega_{RMS}
\\
0  & \Omega_{RMS} &{\delta_p}
\end{array} \hskip-2pt \right)
+ \bar{\delta}\, 
{\Omega_p \Omega_s \over \Omega_{RMS}^2}
 \left(\hskip-2pt \begin{array}{ccc}
{\Omega_p \over \Omega_s}&  -1 &  0
\\
-1  &{\Omega_s \over \Omega_p}& 0
\\
0&0&0
\end{array}\hskip-2pt\right)
\end{equation}
where $\bar{\delta}=\delta - \delta_0$ is the deviation from the trapping condition. 

We use this representation to study detrapping-induced transport in small networks. 
The system is initialized in the trapped state $\ket{D}$ and eventually decays to a sink 
with a rate $\Gamma_{2S}$ \tcred{(see Fig.)}. The sink is accounted for by adding an extra level 
to the network or by adding an imaginary part $-i \Gamma_{2s}/2$ to $\delta_p$. As it is clear from Eq.(\ref{eq:hamil-bright}) a nonzero $\bar{\delta}$ triggers transitions $\ket{D} \to \ket{B}$. As a consequence also $\ket{2}$ populates and decays to the sink. Therefore detrapping is detected by populating the sink. In Fig.\ref{fig:lambda_lightbasis} we show population of the sink $P_{sink}(t) = 1 - \mathrm{Tr}[\rho(t)]$ where $\rho(t)$ is the density matrix of the network, at a large enough time $t=t_f$, for a symmetric network 
$\Omega_s=\Omega_p$ in Lambda configuration $\Omega_0 =0$, where the trapping condition is $\delta_0=0$. In quantum optics the minimum in the curves Eq.(\ref{eq:hamil-bright}) at $\delta=\delta_0$ is referred as the \enquote{dark resonance}.

In the limit $\delta \ll \Omega_{RMS} \lesssim \Gamma_{2S}$ and a simple approximate analytic form may be derived by perturbation theory (see Appendix~\ref{app:current-pert})
\begin{equation}
\label{eq:sinkpop}
\mathrm{Tr}[\rho(t)] \approx \rho_{DD}(t) \approx \exp\Big(
- {\bar{\delta}^2 \over 4 \Omega_{RMS}^2} \Gamma_{2S} \,t
\Big)
\end{equation}
where $\rho_{DD}(t)$ is the population of the trapped state. The formula is derived assuming 
that coupling between the dark and the bright state is small, whereas the coupling between the bright state and site 2 is relatively large as well as the decay to the sink. Decay is triggered by the level splitting $\delta$ and as a result the population of the network decreases. 

The left panel of Fig.\ref{fig:lambda_lightbasis} this result is plotted against the numerical 
calculation, and we see good agreement for range of $|\delta| < |\Omega_{p,s}|$ 
whereas for larger $\delta$ the processes $\ket{D} \to\ket{B}$ and $\ket{B} \to \ket{2}$ 
are no longer separable. In the right panel the effect of changing the decay rate $\Gamma_{2S}$
is shown. Here, the prediction in Eq.~(\ref{eq:sinkpop}) still holds and we see that by increasing 
$\Gamma_{2S}$, the width of the dark resonance is reduced. 

\begin{figure}[t!] 
  \centering 
\includegraphics[width=0.49\textwidth]{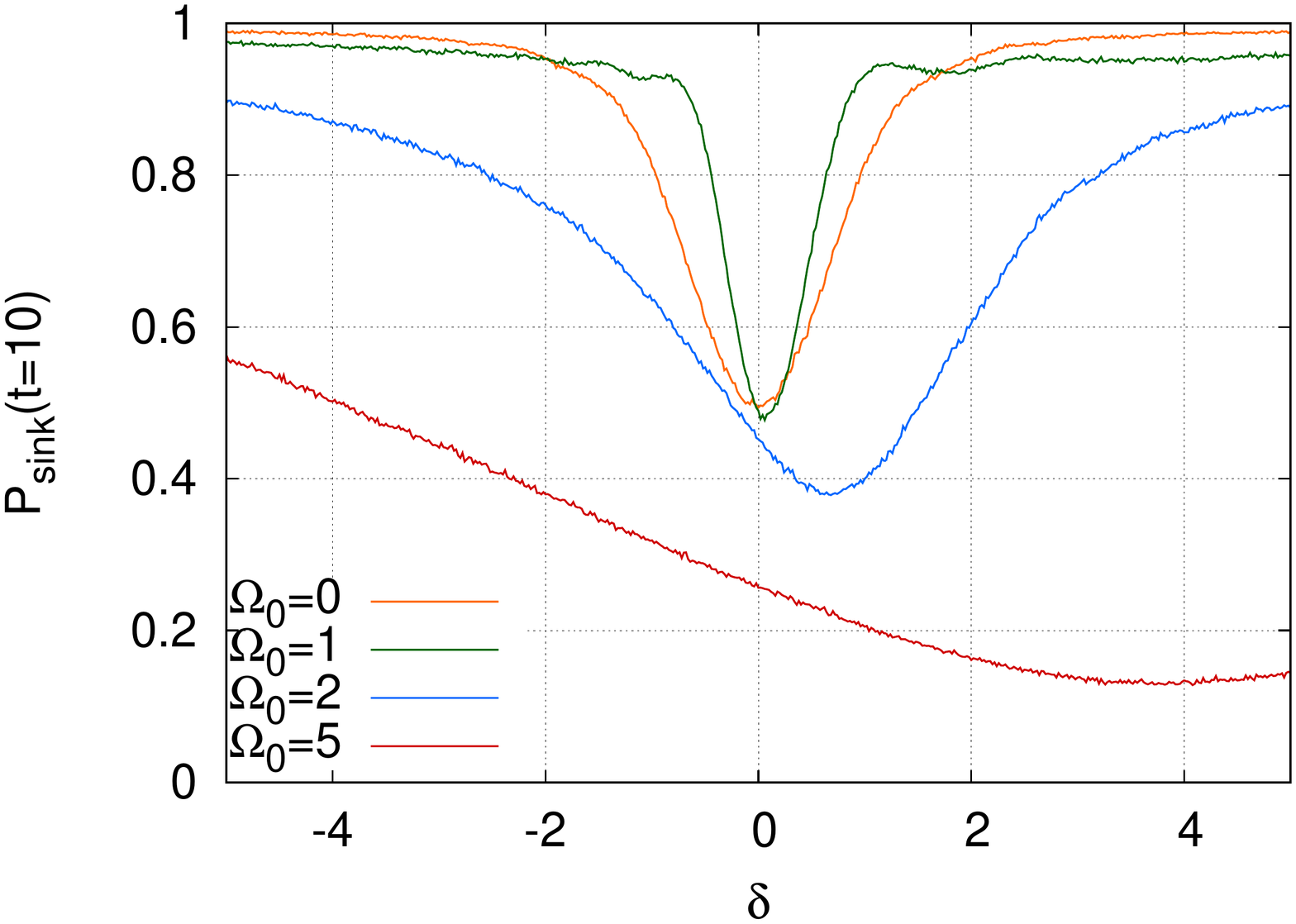}
\includegraphics[width=0.49\textwidth]{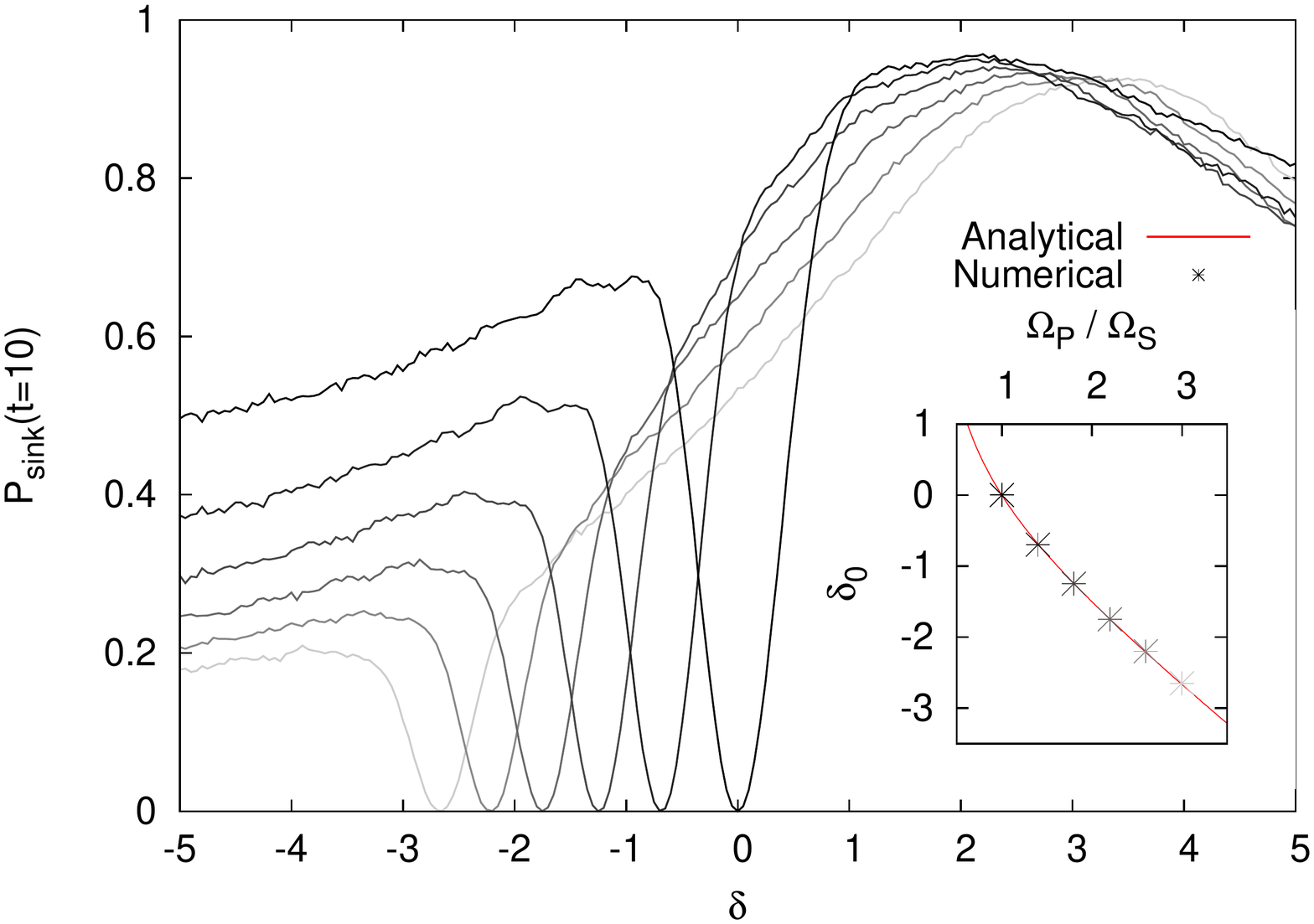}
 \caption{Left0: Numerically calculated final sink population for the $\Lambda$ network for a range of $\delta$ and $\Gamma_{2S}$, where $\Omega=1$ and $\delta_{p}=\Omega_{0}=0$, compared to the analytical prediction for sink. Left: 
Right: Dark state shifting with level splitting $\delta$ in the asymmetric $\Delta$ network. 
  }\label{fig:flucsforin0}
 \end{figure}

The right panel of Fig.\ref{fig:flucsforin0} shows the effect of changing the decay rate $\Gamma_{2S}$. Here, the prediction in Eq. 2.19 still holds and we see that by increasing $\Gamma_{2S}$, the width of the dark resonance is reduced.

Results a Delta system initialized in $\ket{0}$, are shown in the left panel of Fig.(\ref{fig:lambda_lightbasis}). It is apparent that dark resonances in the population appear for finite $\delta$, since the trapping condition  $\delta=\delta_0$
now depends on $\Omega_0$. Notice that population at the minima $\delta=\delta_0$ is now nonzero since only a fraction  
$|\langle 0 | D \rangle|^2= \Omega_s^2/\Omega_{RMS}^2$ of the population is initially trapped. 

 \subsection{Effect of Noise}
In this section we discuss the effect of noise, focusing on the most relevant process destroying coherence, namely dephasing in the trapped state. In addition we consider decay of population from $\ket{2}$ to a sink, described by a fourth state $\ket{S}$. The effect of a Markovian environment is accounted for by the following Lindblad equation
\begin{equation}
\label{eq:lindblad}
\dot{\rho}\left(t\right) =
-\frac{i}{\hbar}\left[H,\rho\left(t\right)\right]+\mathcal{D}_{sink}\left[\rho\left(t\right)\right]+\mathcal{D}_{deph}\left[\rho\left(t\right)\right]
\end{equation}
where decay to the sink and the associated dephasing is described by the dissipator
\begin{equation}
\label{eq:lindblad-sink}
\mathcal{D}_{sink}\left[\rho\left(t\right)\right] := 
\Gamma_{2S} \, \big[ \ket{S}\!\!\bra{S}\,\rho_{22}(t) - 
{1 \over 2} \, \big\{\ket{2}\!\!\bra{2}\,\rho \big\} \big]\; ,
\end{equation}
where $\rho$ is the density matrix of the four state network and curly brackets indicate the anticommutator. Pure dephasing~\cite{kr:14-paladino-rmp} of the trapped state is accounted for by the dissipator
\begin{equation}
\label{eq:lindblad-pure}
\mathcal{D}_{deph}\left[\rho\left(t\right)\right] := \gamma \, \big[\ket{1}\!\!\bra{1}\,\rho_{11}(t) - {1 \over 2} \, \big\{\ket{1}\!\!\bra{1}\,\rho \big\} \big] \; ,
\end{equation}
We also consider detrapping due to non-Markovian calssical noise, which is accounted for by a stochastic process
$x\left(t\right)$ coupled to the Hamiltonian 
\begin{equation}
H \,\to \, H +x\left(t\right)\ket{1}\!\!\bra{1}.
\end{equation}
the evolution of $\rho(t)$ being obtained by averagin over all realizations of $x(t)$. here we consider the Ornstein-Uhlenbeck (OU) classical stochastic process~\cite{kb:85-gardiner-stochastic,ka:01-bartosch-ijmp-ougeneration}, characterized by the un-normalized autocorrelation function; $\braket{x\left(t\right)x\left(t+t^\prime\right)}=\sigma_{ou}^{2}e^{-t^\prime/\tau}$, where $\tau$ is the correlation time. The power spectrum of the noise is given by 
\begin{eqnarray}
\label{eq:powerspec}
\mathcal{S}\left(\omega\right)&= \int_{-\infty}^{\infty}
\hskip-3pt  dt^\prime \; 
e^{i\omega t^\prime}\, 
\braket{x\left(t\right)x\left(t+t^\prime\right)}
\nonumber\\& = 
2\,\sigma_{ou}^{2}\int_{0}^{\infty}
\hskip-3pt  dt \; 
e^{-t/\tau}\,\cos{\left(\omega t\right)}=\frac{2\, \sigma_{ou}^{2}\tau}{1+\omega^{2}\tau^{2}}.
\end{eqnarray}
where we have assumed that the noise is stationary, $\braket{x\left(t\right)x\left(t+t^\prime\right)} = \braket{x\left(0\right)x\left(t^\prime\right)}$. In order to understand the role of non-Markovianity, we will compare noises which would give the same decay time of coherence if treated in by the standard Bloch-Redfield master equation 
theory~\cite{kb:98-cohentann-atomphoton}, since this latter approach yields correct decay rates in the Markovian limit. A scale for the decay time is $\gamma \sim \mathcal{S}\left(\Omega_{0}\right)$, obtained by arguing that local dephasing triggers transitions between  ``dressed'' eigenstates $\ket{\phi_n}$, whose splitting is $\tilde \Omega_{0}$. To this end, for a given $\gamma$ we consider a OU process with width $\sigma_{ou}^{2}$ given by
\begin{equation}
\sigma_{ou}^{2}=\frac{1+\Omega_{0}^{2}\tau^{2}}{\tau}\gamma,
\end{equation}
In the simplest case, where $\Omega_{s}=\Omega_{p}=\tau=1$ and $\Omega_{0}=\delta_{p}=0$ we identify
$\sigma_{ou}^{2}=\gamma$. 

\begin{figure}[!ht] 
\centerline{\includegraphics[width=0.9\textwidth]{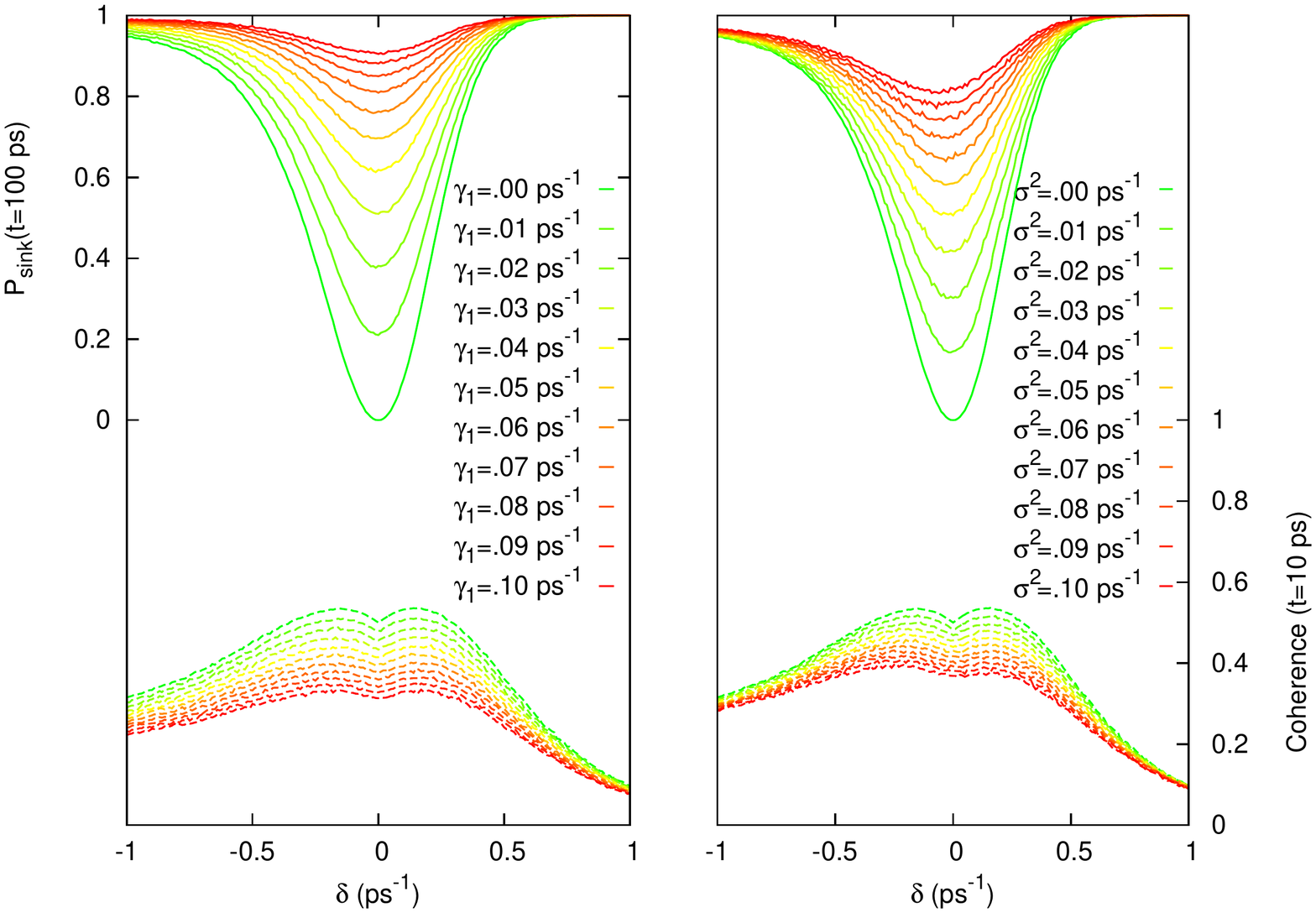}}
\caption{Effects of noise on the $\Delta$ network, where $\Omega_{0}=\delta_{p}=1$. The left-hand panel shows the effect of white noise and the right-hand panel shows the effect of the Ornstein Uhlenbeck process with a correlation time of $\tau=1$ ps. In both panels, the population of the sink at time $t=100$ ps is plotted with solid lines and the sum of the absolute values of the 
coherences at time $t=10$ ps is plotted with dashed line.}\label{deltadarknoise}
\end{figure}

Figure \ref{deltadarknoise} shows the effect of each noise source on the sink population and 
on the sum of the coherences as a function of the level splitting $\delta$ for the $\Delta$ network, where $\Omega_{0}=\delta_{p}=1$. The left-hand panel shows the effect of white noise characterized by a range of $\gamma_{1}$ and the right-hand panel shows the effect of the OU process characterized by a range of noise widths, $\sigma_{ou}^{2}$. The measurement for the coherence is taken at 10 ps, and the measurement for the sink population is taken at 100 ps. It is seen that while decoherence in the  trapped subspace in general weakens the trapping phenomenon an environment with memory is less 
effective in this respect, although in both cases coherences are suppressed in a comparable fashion. 
Non-Markovianity is expected to produce more pronounced signatures if also fluctuations of the splitting $\delta_p$ are accounted for~\cite{ka:213-falci-prb-stirapcpb} together with their correlations, a problem already explited in superconducting quantum architectures~\cite{ka:08-darrigo-njp-crosscorrelations}, where this analysis allows to determine or design optimal operating points where decoherence is minimized~\cite{ka:12-chiarellopaladino-njp-decoherence,ka:10-paladino-prb-opt2qubit,ka:11-paladino-njp-univ2qubit}. 

\begin{figure}[!t] 
\centering 
\includegraphics[width=0.44\textwidth]{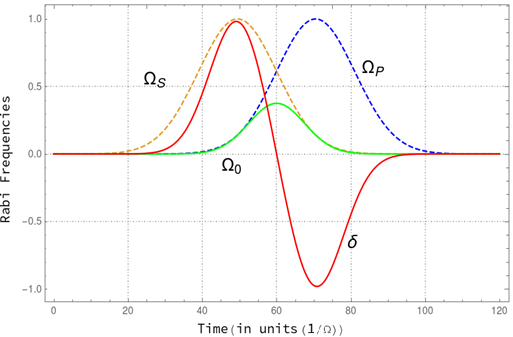}
\qquad
\includegraphics[width=0.44\textwidth]{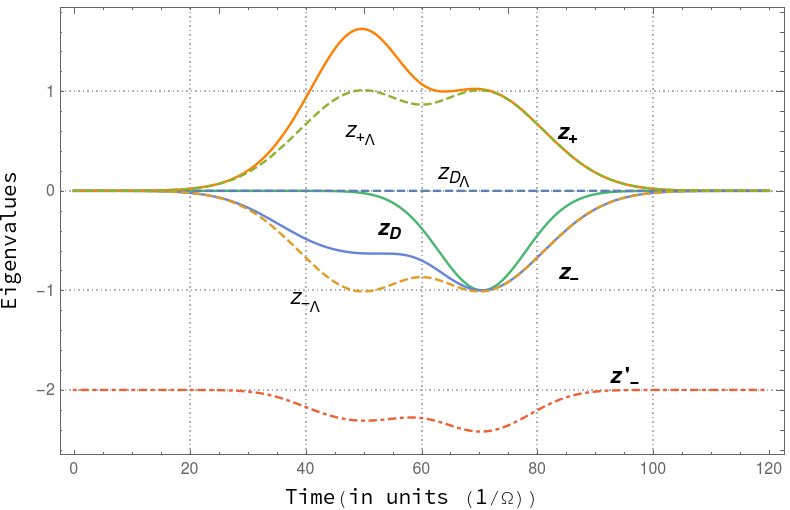}
\\
\includegraphics[width=0.44\textwidth]{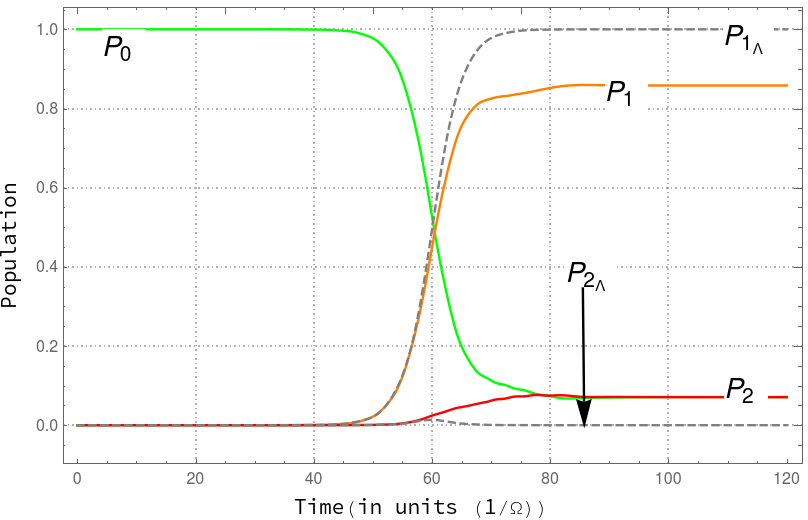}
\qquad
\includegraphics[width=0.44\textwidth]{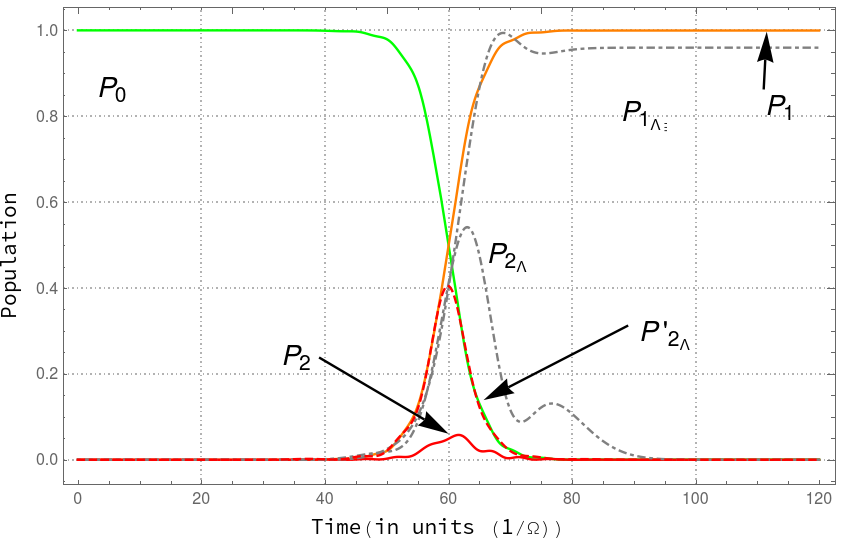}
\caption{Population transfer by STIRAP in the $\Lambda$ ($\alpha=0$ in Eq.~(\ref{eq:omega0})) and in the $\Delta$ ($\alpha=1$) systems for $\Omega T=15$, $\tau=0.7\,T$ and $\gamma=0$. Top left panel: the pulses $\Omega_{s,p}(t)$ for standard STIRAP (dashed lines) and the additional pulses  $\Omega_{0}(t)$ and $\delta(t)$ for unconventional STIRAP (full lines). 
Top right panel: instantaneous eigenvalues for $(\alpha,\delta,\delta_p) =(0,0,0)$ 
(dashed lines) and for $(\alpha,\delta,\delta_p) =(1,\delta_0(t),0)$ (full lines), 
and the eigenvalue $z_-(t)$ for {$(\alpha,\delta,\delta_p) =(1,\delta_0(t),-2 \,\Omega)$}
(dot dashed line), as given by Eqs.(\ref{eq:dark-eigenvalue},\ref{eq:eigenvalues}).
Bottom left panel: population histories for 
for $(\alpha,\delta,\delta_p) =(0,0,0)$ 
(dashed gray lines) and for $(\alpha,\delta,\delta_p) =(1,\delta_0(t),0)$ 
(full lines); these latter showing imperfect population transfer due to Zener tunneling 
occurring when eigenvalues $z_D(t)$ and $z_-(t)$ are degenerate.
Bottom right panel: population histories for 
\tcred{$(\alpha,\delta,\delta_p) =(1,\delta_0(t),-2 \,\Omega)$}
(full lines),  for 
\tcred{$(\alpha,\delta,\delta_p) =(0,\delta_0(t),-2 \,\Omega)$}
(gray dashed lines) and $\rho_{22}(t)$ for $(\alpha,\delta,\delta_p) =(0,0,0)$; 
here the variuos $\rho_{22}(t)$ are magnified by a factor 30, showing that the 
unconventional pattern for STIRAP is more robust against transitions to the intermediale level. 
\label{fig:stirap}}
\end{figure}

\section{Coherent population transfer in Delta systems}
\label{sec:stirap}
Adiabatically modulated trapped states can be used for implementing quantum operation, the  
simplest being is population transfer~\cite{kr:17-vitanovbergmann-rmp}. This can be achieved by suitable modulation of the parameters
of the Hamiltonian (\ref{eq:hamiltonian}). Physically it describes a three-level atom driven by external fields $W_k(t)=  2 \Omega_k \, \cos(\omega_k t)$, for $k=0,s,p$. The angular frequencies 
$\omega_i$ of the fields are chosen to approximately match the level splittings of the atom, therefore they address specific transitions. In particular we take $\omega_p = \epsilon_2 - \epsilon_0 -\delta_p$, where $|\delta_p| \ll \omega_p$ is the pump field detuning. We also take slightly detuned, $\omega_s = \epsilon_2 - \epsilon_1 -\delta_s$, whereas $\omega_0= \epsilon_1 - \epsilon_0$ is not detuned.
We define the two-photon detuning $\delta=\delta_p-\delta_s$. Then Eq.(\ref{eq:hamiltonian}) is the 
Hamiltonian in the rotating wave approximation as seen from a rotating frame. 

\begin{figure}[!t] 
\centering 
\includegraphics[width=0.44\textwidth]{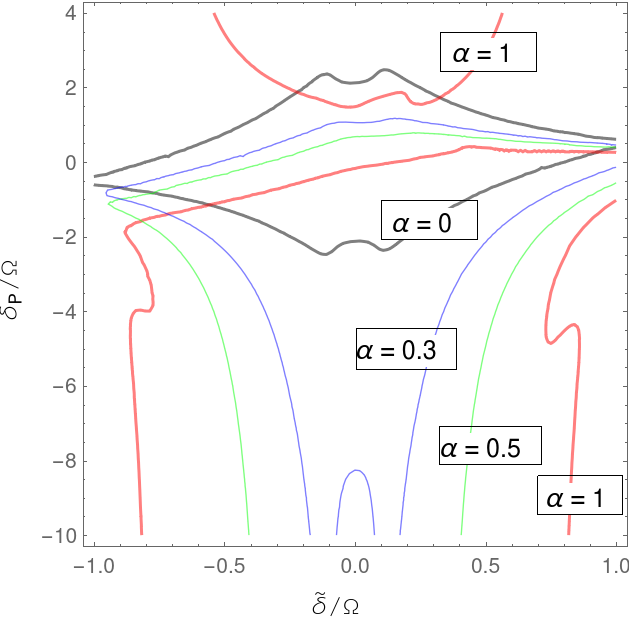}
 \caption{Stability against fluctuations of detunings of population transfer by STIRAP,
for  $\Omega T=15$, $\tau=0.7\,T$  and $\gamma=1/T$ and various $\alpha=0,0.3,0.5,1.0$. 
We plot lines corresponding to $P_1(t_f)=0.95$. The stability region enlarges for increasing 
$\alpha$ showing that that the $\Delta$ system protocol is more robust. A negative $\delta_p$ 
increases the stability for larger $\alpha$ since it removes the degeneracy  
of $z_D(t)$ and $z_-(t)$ enforcing the adiabatic pattern $\ket{0} \to \ket{1}$. Also 
a positive $\delta_p$ may increase population transfer efficiency, due to pattens combining 
adiabatic dynamics an diabatic transitions.}\label{fig:stirap-stability}
\end{figure}

We allow for a time dependence of the Rabi frequencies $\Omega_k(t)$ on a time scale much longer than the field period $2 \pi /\omega_k$. Therefore  the Hamiltonian (\ref{eq:hamiltonian}) may determine an  adiabatic dynamics of the system. In the $\Lambda$ configuration at two-photon resonance, 
$\delta=0$ this allows to achieve complete population transfer $\ket{0} \to \ket{1}$ by adiabatically following the trapped state Eq.(\ref{eq:trapped}). To this end the fields' amplitudes are adiabatically modulated  in the \enquote{counterintuitive} sequence, i.e. $\Omega_s(t)$ is shined before $\Omega_p(t)$, a protocol named STIRAP. 
In our work we consider for simplicity Gaussian pulses of equal amplitude
$$
\Omega_s(t) = \Omega \, \mathrm{e}^{- \left({t+\tau \over T}\right)^2}
\quad ; \quad
\Omega_p(t) = \Omega \, \mathrm{e}^{- \left({t-\tau \over T}\right)^2}
$$
where $T$ is the width of the pulse and the delay is $\tau \lesssim T$, 
shown in Fig.~\ref{fig:stirap}. The protocol is very selective and robust, being almost insensitive to fluctuation of all parameters, except $\delta$ \tcred{(see Fig.~\ref{fig:stirap-stability})}.  
Population transfer by STIRAP~\cite{kr:17-vitanovbergmann-rmp} in solid-state devices has been proposed in systems of artificial atoms based on superconductors~\cite{ka:06-siebrafalci-optcomm-stirap,ka:205-liunori-prl-adiabaticpassage,ka:09-siebrafalci-prb,ka:17-falci-fortphys-fqmt} and semiconductors~\cite{kr:16-menchongreentree-repprogphys-spatialadpass} and has been recently demonstated in Josephson systems~\cite{ka:16-kumarparaoanu-natcomm-stirap,ka:16-xuzhao-natcomm-stirapexp}.

STIRAP can also be implemented in the $\Delta$ system, as shown in Fig.~\ref{fig:stirap-stability} (bottom panels). We have chosen the following form for the coupling 
\begin{equation}
\label{eq:omega0}
\Omega_0(t) = \alpha  \,{\Omega_s(t)\,\Omega_p(t) \over \Omega} =  
\alpha  \,\Omega\, \mathrm{e}^{- {2 \over T^2} (t^2+\tau^2)}
\end{equation}
population transfer being achieved by modulation of the two-photon detuning $\delta(t)= \delta_0(t)$ where from Eq.(\ref{eq:dark-condition}) 
$$
\delta_0(t)= {\Omega_0(t) \over \Omega_s(t) \Omega_p(t) } \, [\Omega_s^2(t) - \Omega_p^2(t)] =
{\alpha \over \Omega} \, [\Omega_s^2(t) - \Omega_p^2(t)]
$$ 
The modulated fields are shown in Fig.~\ref{fig:stirap}. Also the instantaneous eigenvalues are plotted, showing that an \textquote{unconventional} adiabatic pattern for population transfer exists also in the Delta system with modulated $\delta(t)$. It is interesting to notice that for $\delta_p=0$ instantaneous eigenstates may be degenerate at some time ($z_D$ and $z_-$ in the top right panel of Fig.~\ref{fig:stirap}), therefore efficiency of the protocol is lowered by Zener tunneling (see Fig.~\ref{fig:stirap}, lower left panel). Efficiency is recovered by using a constant $\delta_p <0$ (see Fig.~\ref{fig:stirap}, lower right panel) which eliminates the crossing. In this case not only the final population of the target state is larger, but population of the intermediate state (which in many cases should be avoided) is lowered enforcing trapping during the whole protocol. 

Lower population of the intermediate state is a fingerprint of the fact that \enquote{unconventional} 
STIRAP in $\Delta$ configuration is more robust than the conventional $\Lambda$-STIRAP. 
This is clearly shown in Fig.~\ref{fig:stirap-stability}, where we study sensitivity of the efficiency $P_1(t_f)$ to parametric fluctuations of the detunings. Thus we allow for constant deviations of the detunings from their reference values, namely $\delta(t) = \delta_0(t) + \bar{\delta}$  and $\delta_p = \bar{\delta}_p$. 
In order to check that population is trapped during the whole procedure, we introduce decay of the intermediate level $\delta_p \to \delta_p - i \gamma/2$ which lowers $P_1(t_f)$ if $\ket{2}$ is always occupied. As it is apparent the presence of the field $\Omega_0$ enlarges the stability region. This latter is moved towards negative values of $\bar{\delta}_p$ which ensure that crossings of adiabatic eigenvalues are eliminated. Actually efficiency may increase also for $\bar{\delta_p}  >0$ (effect not shown in the figure), due to the combination of adiabatic evolution and diabatic transition where Zener tunneling restores the pattern leading to population transfer. An analogous phenomenon is known to occurs in $\Lambda$ STIRAP for $\delta$ nonvanishing~\cite{kr:17-vitanovbergmann-rmp,ka:12-falci-physscripta-drivencoherent}, which in general guarantees population transfer also for $\delta$ sufficiently close to the trapping condition
Eq.(\ref{eq:dark-condition}). 

A $\Delta$ network can be implemented in superconducting architectures by working away from symmetry points~\cite{kr:11-younori-nature-artificialatoms} but decoherence is minimized at optimal points, where one of the three couplings has to be implemented by a two-photon process~\cite{ka:16-distefano-pra-twoplusone}. Other protocols combining slow modulation of detuning with resonant ac pulses have also been proposed~\cite{ka:15-distefano-prb-cstirap}, which in artificial atoms are expected to be robust against phase noise since control is operated at microwave frequencies.  

\section{Conclusions}
We have studied coherent trapping in three-site quantum networks. We have shown that 
under suitable conditions $\Delta$ networks have trapped eigenstates analogous to   
what is found in $\Lambda$ (and also Ladder and Vee) networks. 

Trapping and detrapping phenomena could be relevant in several physical systems, as 
artificial networks of semiconducting quantum dots~\cite{ka:14-plenio-njp-tripleqdot} or molecular complexes in certain biological systems~\cite{kb:14-mohseni-qbiol}, where the $\Delta$ system describes either part or the whole network, and for small circuit-QED quantum networks of artificial atoms where the $\delta$ system may provide an effective low-energy description. 

In driven $\Delta$ network the physics of trapping may allow adiabatic coherent population transfer.
Delta systems in superconductors have been studied~\cite{kr:11-younori-nature-artificialatoms} and recently implemented to demonstrate superadiabatic protocols. This configuration of couplings may also be relevant for adiabatic dynamics in circuit-QED architectures~\cite{ka:204-wallraff-superqubit,ka:18-distefanopaternostro-prb-measurethermo}. 

The rich physics of the dark resonance in $\Delta$ systems can be used as a tool for sensing the properties of a noisy environment~\cite{ka:08-darrigo-njp-crosscorrelations} yielding for instance direct information on dephasing and on its non-Markovianity~\cite{kr:14-paladino-rmp} or to find optimal operating point in quantum devices~\cite{ka:12-chiarellopaladino-njp-decoherence,ka:10-paladino-prb-opt2qubit,ka:11-paladino-njp-univ2qubit}. 
\appendix 
\section{Derivation of the current in the sink}
\label{app:current-pert}
To derive Eq.(\ref{eq:sinkpop}) we use perturbation theory in the two-photon detuning $\bar{\delta}$, by writing the Hamiltonian Eq.(\ref{eq:hamiltonian}) using the eigenbasis for $\delta=\delta_0$  
$$
H = \sum_{k=D,\pm} z_k \,\ket{k}\!\!\bra{k} + \bar{\delta} \,  \,\ket{1}\!\!\bra{1} 
$$ 
In the limit $\delta \ll \Omega_{RMS} \lesssim \Gamma_{2S}$ we can suppose that $\ket{\pm}$ is 
almost depopulated and $\mathrm{Tr}[\rho(t)] \approx |\langle {D| \psi(t)}\rangle|^2$. 
We prepare the system in $\ket{D}$. Then for small $\bar{\delta}$ in leading order in 
$\bar{\delta}/\Omega_{RMS}$ the only nonvanishing amplitude is $\langle {D| \psi(t)}\rangle \approx 1$ and 
$$
\langle {D| \psi(t)}\rangle \approx \mathrm{e}^{- i (z_D + \delta z_D) t} 
$$ 
where $\delta z_D$ is the correction to the eigenvalue $z_D$. In the illustrative case 
of the symmetric $\Lambda$ system, $\Omega_0=0$ and $\Omega_p=\Omega_s= \Omega_{RMS}/\sqrt{2}$ 
using the eigenstates (\ref{eq:trapped},\ref{eq:eigenvectors}) and the eigenvalues 
(\ref{eq:dark-eigenvalue},\ref{eq:eigenvalues}) we easily evaluate 
$$
\delta z_D = {\bar{\delta} \over 2} - {\bar{\delta}^2 \over 4} \,\Big[ {\sin^2 \Phi \over z_+} 
+ {\cos^2 \Phi \over z_-}  \Big] =  
{\bar{\delta} \over 2} - {\bar{\delta}^2 \delta_p \over 4 \Omega^2_{RMS}} 
$$
Finally letting $\delta_p \to \delta_p - i \Gamma_{2S}/2$, we obtain 
$$
|\langle {D| \psi(t)}\rangle| \approx \mathrm{e}^{2 \,\Im (\delta z_D) \,t} = 
\mathrm{e}^{- {\bar{\delta}^2 \Gamma_{2S} \over 4 \Omega^2_{RMS}} t}
$$ 
which is the desired result. Eq.(\ref{eq:sinkpop}) may be derived more rigorously and extended beyond the small $\bar{\delta}$ by adiabatic elimination of the bright state. 

\providecommand{\newblock}{}


\end{document}